\documentclass[intlimits,twoside,a4paper]{article}

\usepackage[cp1251]{inputenc}
\usepackage{subfigure}
     
   \usepackage{float}

\usepackage[eqsecnum]{cmpj3}

\issue{2026}{29}{2}{23702}
\doinumber{10.5488/CMP.29.23702}

\title[First-principles study of the impact of as doping]
{First-principles study of the impact of as doping on the structural and electronic properties of MoS$_2$ monolayer}
\author[A. Daouadi, M. L. Benkhedir]{A. Daouadi\orcid{0009-0008-6640-5656}, 
	M. L. Benkhedir\orcid{0000-0001-8375-0998}\thanks{Corresponding author:\email{mohamed.benkhedir@univ-tebessa.dz}.}}
   
\address{Laboratory of Theoretical and Applied Physics (LPAT), Echahid Cheick Larbi Tebessi University, 12000 Tebessa, Algeria}


\date{Received 5 July 2025; revised 31 December 2025; accepted 30 January 2026; published 29 June 2026}

\begin{document}
	
 \maketitle
	
\begin{abstract}	
	This study is aimed at exploring the structural and electronic properties of doped MoS$_2$ monolayers, including Mo and S vacancies and As doped systems, employing DFT calculations. The electronic properties were analyzed to understand how these modifications affect the behavior of the material. Introduction of defects generates new defect states in the midgap. In the S-vacancy (V$_\text{S}$), Mo- vacancy (V$_{\text{Mo}}$), As-Mo (As substituting Mo), and As-S (As substituting S) doped systems, the downward shift of the Fermi level to the valence band indicates a $p$-type behavior. In the As interstitial system the Fermi level shifts to the conduction band, suggesting an $n$-type semiconductor. The results highlight that doping MoS$_2$ with As, particularly at the Mo site, can be used in photocatalysis and high-efficiency photovoltaics. Additionally, the As interstitial system demonstrates an enhanced performance in field-effect transistors (FETs).

	%
	\printkeywords
\end{abstract}

\section{Introduction}

The discovery of graphene triggered a global surge in research on two-dimensional (2D) materials due to their exceptional physical, chemical and electronic properties \cite{novoselov2004electric}. Graphene, a monolayer of carbon atoms formed in an hexagonal structure, exhibits a combination of exceptional characteristics \cite{han2007energy, lee2008measurement}. The success of graphene not only earned the Nobel Prize in Physics in 2010 \cite{RevModPhys.81.109} but also opened the door to an entirely new class of materials called two-dimensional (2D) materials.  Transition metal dichalcogenides (TMDs) in  their 2D form demonstrate a wide variety  of properties, including electronic, mechanical, optical and chemical. In contrast to gapless graphene, TMDs have notable band gaps, making them highly effective for applications in field-effect transistors (FETs) and optoelectronics  \cite{wilson1969transition,wang2012electronics,das2014tunable}. The general configuration of these materials is MX$_2$, where M symbolizes  transition metal  and the chalcogen atom is denoted by X as seen in compounds like in MoS$_2$, MoTe$_2$, NbS$_2$, NbSe$_2$, WS$_2$, TaS$_2$, MoSe$_2$, WSe$_2$. The crystal structure of TMDs generally comprises three atomic layers organized in a sandwich (X-M-X), with the transition metal atom layer positioned between two chalcogen layers. These structural units are bounded  in the bulk form by van der Waals forces, while the atoms in-plane are bonded covalently \cite{C4CS00276H}.

As a key member of the TMDs,  molybdenum disulfide (MoS$_2$)  exhibits a distinctive electronic structure that is highly dependent on the number of layers. In its bulk phase, it demonstrates an indirect band gap energy of around 1.2 eV, which limits its applicability in optoelectronic devices. However, when reduced to a monolayer, the material transits to a direct  gap energy  1.8 eV \cite{ding2011first}, that significantly enhances its optical and electronic properties. On the other hand, monolayer MoS$_2$ is rarely free from defects. Point defects, including vacancies and dopants, perform a critical role in modulating its electronic and  optical properties. Recent studies have shown that defects like  sulfur and molybdenum vacancies in MoS$_2$ can have profound effects on the electronic structure and optical absorption~\cite{yang2019electrical, feng2018influence}. In addition to defects, doping has been widely studied as a strategy to modify and enhance the properties of MoS$_2$. Feng et~al.~\cite{feng2014growth} observed that replacing a sulfur atom with selenium in MoS$_2$ monolayer led to a 7\% decrease in the band gap. Another investigation demonstrates that Re and Nb  \cite{noh2015deep} doping generates shallow levels in MoS$_2$ monolayer, facilitating an improved light absorption.  Moreover, introducing transition metals doping such as Mn, Fe and Zn \cite{cheng2013prediction} and also doping with non-metals  like H, B, C, N and F \cite{yue2013functionalization} is a widely explored technique to further enhance the properties of the materials. The doping process can induce $n$-type or $p$-type conductivity, room-temperature ferromagnetism and even strain-dependent modification to the band gap of the materials \cite{tsai2018impact}. Consequently, doped MoS$_2$ has shown enhanced performance in various applications, specifically in catalytic systems for hydrogen evolution and CO$_2$ reduction reactions \cite{balan2023mos2}, field-effect transistors (FETs) with improved on/off ratios~\cite{tong2015advances}.  In optoelectronics, engineered defects enhance light absorption and photocarrier, making MoS$_2$ suitable for high-performance  photodetectors and solar cells~\cite{lopez2013ultrasensitive} and chemical sensors~\cite{perkins2013chemical}.

As shown in several studies~\cite{dolui2013possible}, the properties of MoS$_2$ can be significantly modified by defects. However, to the best of our knowledge there is a lack of studies examining the effect of arsenic (As) either as a substitutional dopant for a Mo and S atoms, or as an interstitial defect. Through  DFT calculations, we investigate how such defects  affect the behavior of  monolayer MoS$_2$. We specifically concentrate on the impact of doping with As doping. Arsenic  is an interesting dopant due to its capability to modify the electronic properties of MoS$_2$ monolayer.  Our research seeks to provide a deeper understanding of  how As doping (substitution and interstitial) specifically alters the electronic properties of MoS$_2$, potentially enhancing its performance for a range of electronic applications.

\section{Details of calculations}

The electronic  properties in all calculation are performed using the PWscf  (Plane-Wave Self-Consistent Field) and ultra-soft pseudepotentials (USPP)  methods as implemented in the  open-source software package Quantum ESPRESSO (QE) which is based on density functional theory (DFT)  \cite{giannozzi2009quantum,giannozzi2017advanced}.
Kohn-Sham density functional theory \cite{baerends2001perspective} is widely
used for self-consistent-field electronic structure calculations. Perdew–Burke–Ernzerhof (PBE) characterization of the generalized-gradient approximation (GGA) was used for modelling  the ionic and electronic interactions \cite{perdew1996generalized,perdew2008restoring}.

Bulk structure of MoS$_2$ consists of layered hexagonal planes (honeycomb), where each layer is made up of a sandwich like configuration of sulfur and molybdenum atoms (S-Mo-S). These layers are weakly bonded by van der Waals forces along the  $z$-axis, which makes it feasible to isolate a single monolayer~\cite{raybaud2000ab}, which has a hexagonal structure with space group P-6m2  (187) \cite{wilson1969transition}. 

In this study, a single S-Mo-S monolayer was extracted by removing the repeated layer in the $z$-axis. To avert the interactions between periodic slabs, an insertion of a vacuum layer about 16 {\AA} was employed. After isolating the monolayer, a supercell was constructed by multiplying the unit cell three times along both the x and y axis, while keeping a single layer in the $z$-axis, forming a   $3\times3\times1$  monolayer supercell. This resulted in a structure containing 27 atoms: 9 molybdenum (Mo) atoms and 18 sulfur (S) atoms. This supercell was used as the basis for further simulations in QE, where the geometry was relaxed  to ensure structural stability before calculating electronic properties [figure~\ref{fig:supercell_structures} (a)]. For computations, we used kinetic-energy cutoff (ecutwfc)  of 50 Ry and an  charge density cutoff (ecutrho) of  440 Ry. A $14\times14\times1$  Monkhorst-Pack k-point mesh  was used to calculate the electronic properties.
To model the different structures in monolayer MoS$_2$. A  $3\times3\times1$  was used as base structure. Vacancy defects were introduced by taking out a single Mo atom to create a Mo vacancy  figure~\ref{fig:supercell_structures} (b), or a single S atom to create a S vacancy  [figure~\ref{fig:supercell_structures} (c)]. For substitution doping, one Mo atom was replaced with As atom [figure~\ref{fig:supercell_structures} (d)] and similarly, substituting a single S atom with an As atom [figure~\ref{fig:supercell_structures} (e)]. Additionally, an interstitial doping model was generated by inserting an As atom [figure~\ref{fig:supercell_structures} (f)]. All defective and doped structures were subjected to the same structural optimization procedure described previously for the pristine MoS$_2$ monolayer, using identical computational parameters to ensure consistency and comparability.

\section{Formation energy}

The formation energy ($E_f$) of a defective supercell is calculated using the equation below \cite{wang2017defects}:
\begin{equation}
E_f = E_{\text{defect}} - E_{\text{perfect}} - \sum_i n_i \mu_i,
\end{equation}
where  $E_{\text{defect}}$ and $E_{\text{perfect}}$ refer to the total energies of defective MoS$_2$  and pristine MoS$_2$ supercell. $n_i$  represents the number of atom $i$ that have been added or removed from the system. If $n_i$ is negative, it means that the atom has been removed. If it is positive, it means the atom has been added.  $\mu_i$ is the  chemical potential corresponding to element $i$, respectively. The chemical potentials are calculated using the bulk rhombohedral phase of As and the bulk body-centered cubic (bcc) phases of Mo and S \cite{bollinger2003atomic}.

The formation energies of all the  doping variants in MoS$_2$ are presented in table~\ref{t}. The results show that the As-Mo system has the minimum formation energy of 2.81 eV compared to the other defect configurations, indicating  its relative stability and  making it more likely to form.  The As-S doped system with a formation energy of 3.49 eV is slightly less stable compared to the As-Mo doped system. By contrast, the As interstitial system has the maximum formation energy of 26.03~eV, making it unstable and difficult to form.

\section{Result and discussion}

\subsection{Crystal structure}

\begin{figure}[H]
	\centering
	\subfigure[]{
		\includegraphics[width=0.39\textwidth]{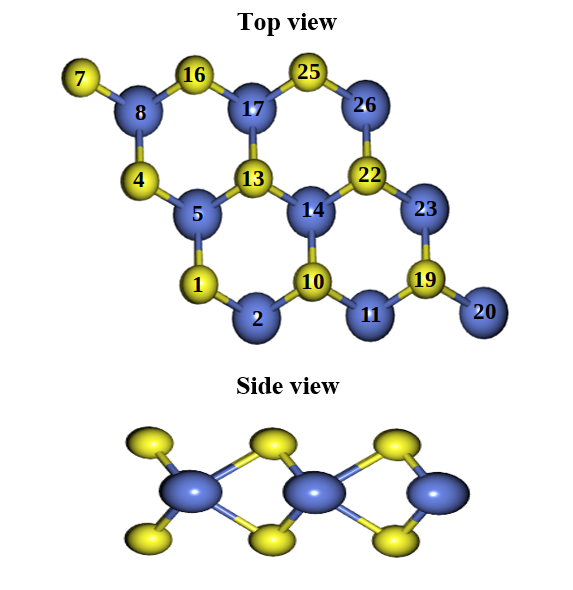}
	}
	\subfigure[]{
		\includegraphics[width=0.33\textwidth]{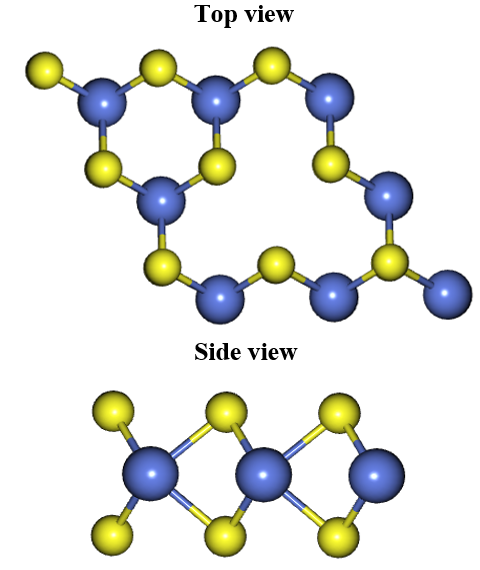}
	}
	\subfigure[]{
		\includegraphics[width=0.35\textwidth]{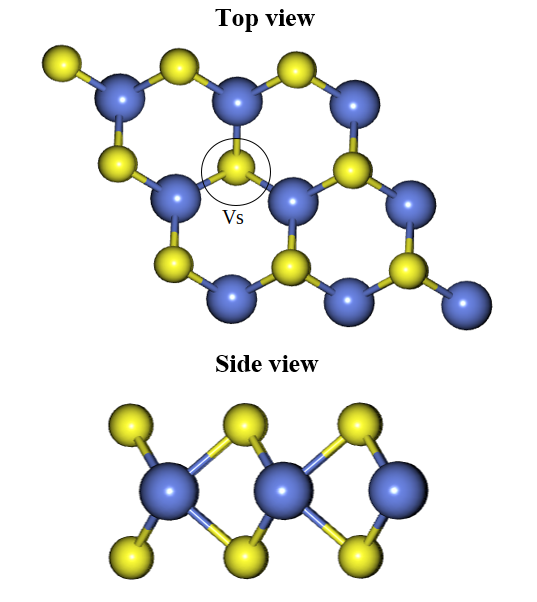}
	}
	\subfigure[]{
		\includegraphics[width=0.33\textwidth]{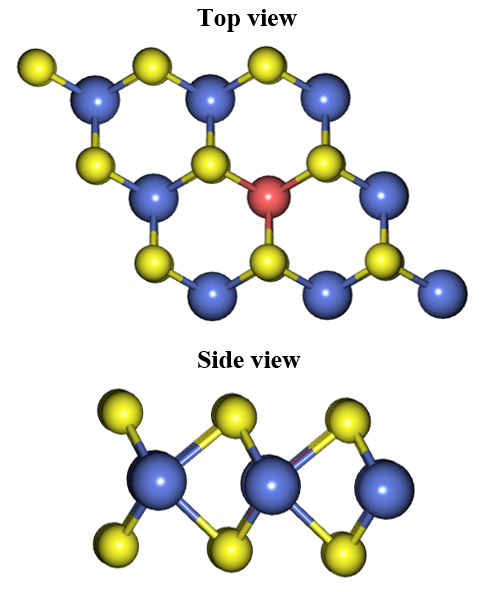}
	}
	\subfigure[]{
		\includegraphics[width=0.33\textwidth]{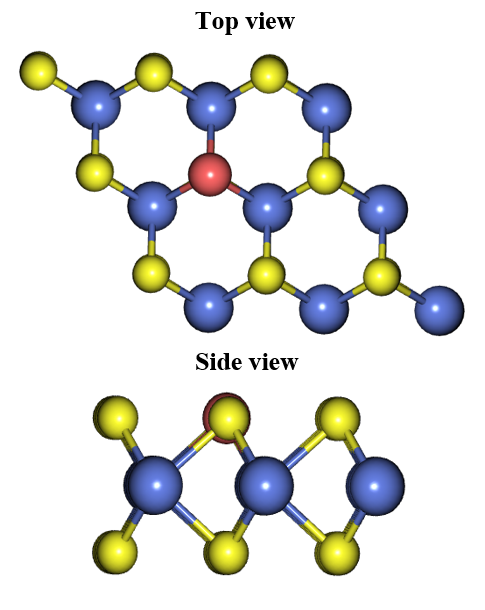}
	}
	\subfigure[]{
		\includegraphics[width=0.34\textwidth]{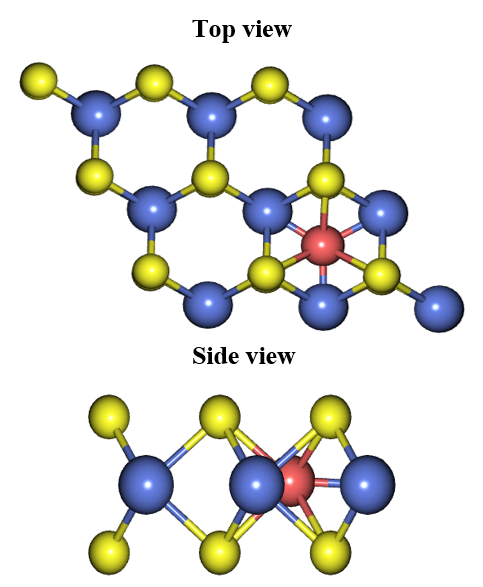}
	}
	\caption{(Colour online) Top and side views of geometric structures of $3\times3\times1$ supercell with 27 atoms: (a) undoped MoS$_2$, (b) V$_{\text{Mo}}$ system, (c) V$_\text{S}$ system,  (d) As-Mo doped system, (e) As-S doped system and (f) As-interstitial system. blue, yellow and red are Mo, S, and As atoms.}
	\label{fig:supercell_structures}
\end{figure}

In our study, the bulk MoS$_2$ was optimised and the obtained  cell perameters are $a = b = 3.17$ {\AA} and $c = 13.22$ {\AA}.  The results align well with the experimental values of $a_{\exp}$ = 3.16 {\AA}, $c_{\exp}$ = 12.29  {\AA} \cite{wilson1969transition} and theoretical values  $a_{\text{theo}}$ = 3.18 {\AA}, $c_{\text{theo}}$ = 13.82 {\AA} \cite{johari2011tunable}, respectively. In the case of Mo-vacancy and S-vacancy, we have taken out the Mo atom number 14 [figure~\ref{fig:supercell_structures} (b)] and the S atom number 13 [figure~\ref{fig:supercell_structures} (c)] for MoS$_2$ monolayer. Subsequently, we substituted the Mo [figure~\ref{fig:supercell_structures} (d)] and S [figure~\ref{fig:supercell_structures} (e)] atoms  with an As atom at each of their respective sites. Furthermore, we introduced an additional As atom into an interstitial site within the monolayer MoS$_2$ [figure~\ref{fig:supercell_structures} (f)].  The introduction  of a vacancy or an  As dopant causes only local atomic distortions without altering the overall lattice symmetry. This is due to the low defect concentration and the preservation of periodic boundary conditions in the supercell. Several other studies observed the same behavior \cite{komsa2015native, zhou2013intrinsic, dolui2013ab, zhong2019electronic}.

\begin{table}[h]
	\caption{Calculated values of bond length Mo-X ({\AA}), formation energy E$_f$~(eV), doping properties and the band gap E$_g$ (eV).}
	\label{t}
	\vspace{2ex}
	\begin{center}
		\renewcommand{\arraystretch}{1.5}
		\begin{tabular}{|c|c|c|c|c|c|}
			\hline	
			\textbf{System}    &  	\textbf{{d$_{\text{Mo-X}}$} ({\AA})}   &   \textbf{Doping properties}   &    \textbf{Formation energy {E$_f$}(eV)}   &    \textbf{Band gap {E$_g$} (eV)}  \strut\\
			
			\hline
			\textbf{Undoped MoS$_2$} &2.41  &semiconductor &/&1.72   \strut\\
			\hline
			\textbf{V$_{\text{Mo}}$}& /& $p$-type& 4.08& 1.9    \strut\\
			\hline
			\textbf{V$_\text{S}$}& /& $p$-type& 3.88&  1.8   \strut\\
			\hline
			\textbf{As-Mo}&  3.24& $p$-type&  2.81& 1.59   \strut\\
			\hline
			\textbf{As-S}& 3.24& $p$-type& 3.49& 1.6  \strut\\
			\hline
			\textbf{As interstitial}&   1.88& $n$-type& 26.03& 1.85  \strut\\
			\hline
		\end{tabular}
	\end{center}
\end{table}

\subsection{Electronic structure}

\begin{figure}[!h]
		\centering
		\includegraphics[width=0.8\textwidth]{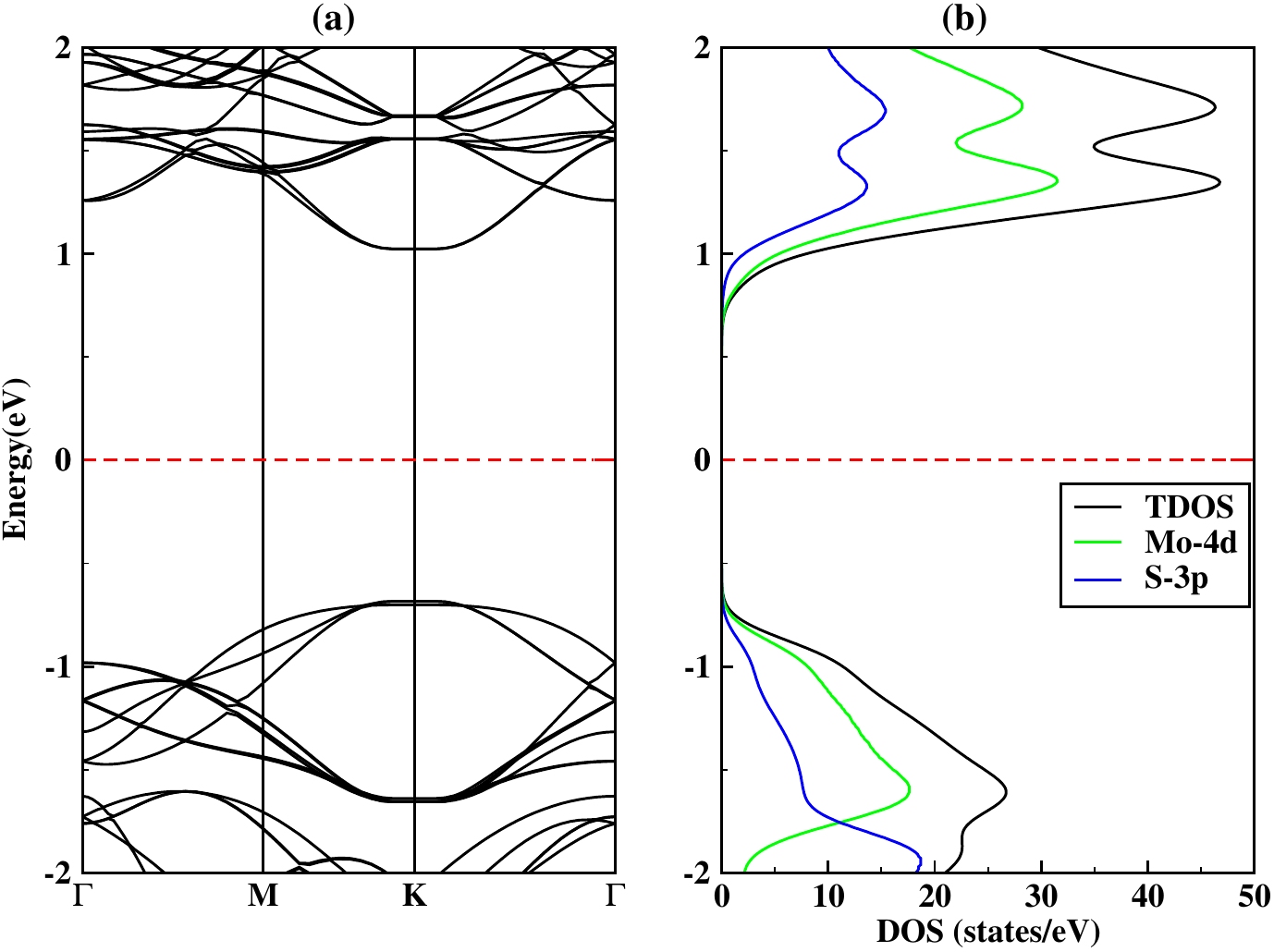}
		\caption{(Colour online) Band structure ($\mathbf{a}$), total and partial DOS of undoped  monolayer  MoS$_2$  ($\mathbf{b}$). Fermi level is set at 0 eV.}
		\label{fig:undoped}
	\end{figure}

The electronic band structure and density of states (TDOS and PDOS) for the undoped MoS$_2$ are presented in figure~\ref{fig:undoped} over an energy range from $-2$ eV to $+2$ eV . The band structure of undoped MoS$_2$  is a direct band gap  semiconductor with the conduction band minimum (CBM) and valence band maximum (VBM) located at the K point of 1.72 eV  showing  good agreement with earlier studies~\cite{ding2011first}. The PDOS  results for undoped MoS$_2$ are displayed in figure~\ref{fig:undoped}~(b). The conduction and valence bands of undoped MoS$_2$ indicate that the contributions came from the  S-3$p$ and  Mo-4$d$  orbitals and a strong hybridization was observed  between them. The absence of any band crossing the Fermi level, which is located in the midgap 
\noindent further confirms the semiconductor nature of MoS$_2$ \cite{lebegue2009electronic}.

The difference between the band gap values from the band structure and DOS mainly results from an insufficiently dense $k$-mesh \cite{setyawan2010high} and a large smearing parameter that blurs DOS features and reduces the apparent gap \cite{toriyama2021comparison}, which tends to blur the sharp features of the DOS by introducing a tail inside the gap, thereby reducing its apparent value. However, this difference does not affect the existence of defects or the relative position of the Fermi level, The same reasoning applies to all the presented systems (Mo-vacancy, S-vacancy, As-Mo, As-S, and inter-As). 

\begin{figure}[h]
	\centering
	\includegraphics[width=0.7\textwidth]{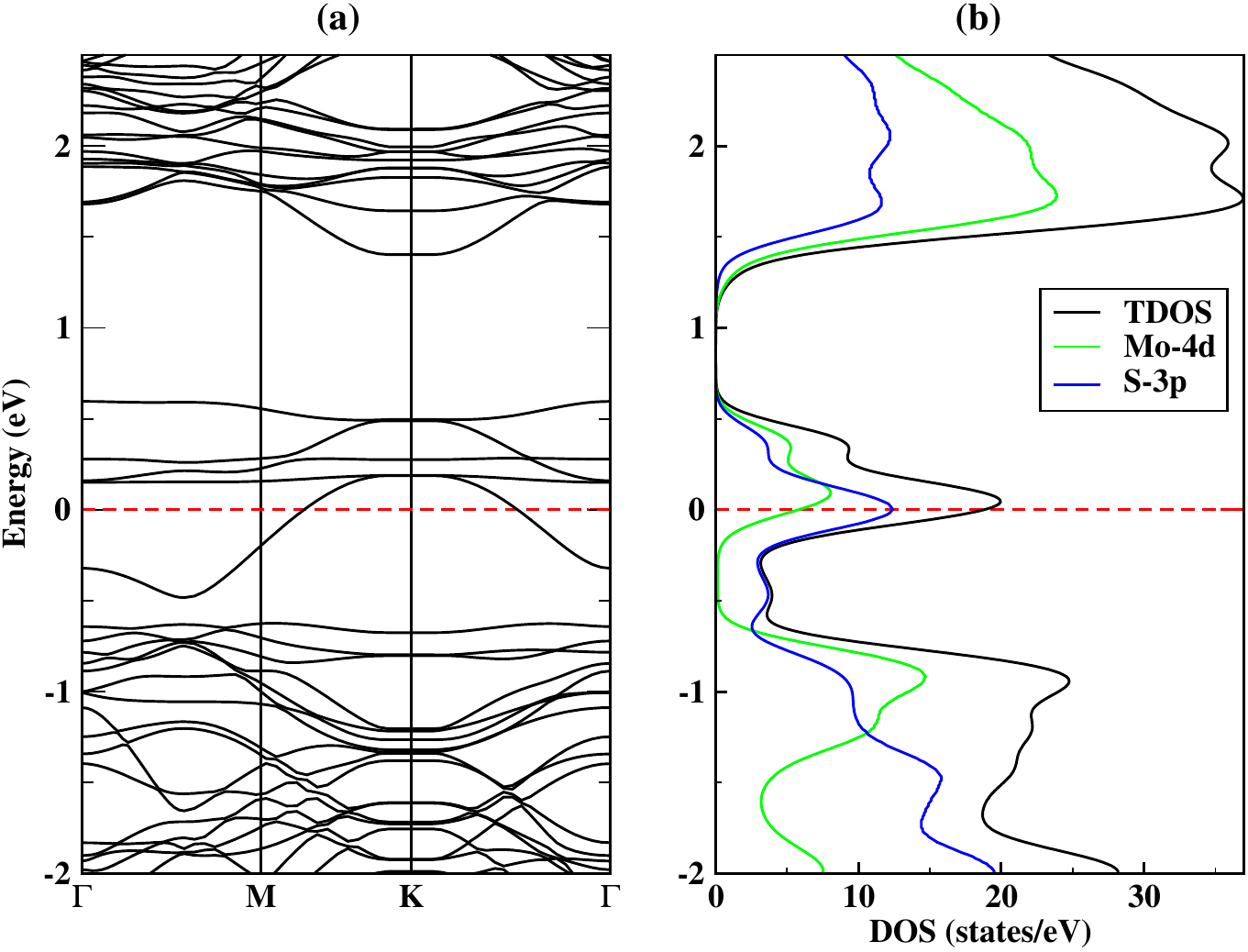}
	\caption{(Colour online) Band structure ($\mathbf{a}$), total and partial DOS of Mo-vacancy in monolayer  MoS$_2$ ($\mathbf{b}$). Fermi level is set at 0 eV.}
	\label{fig:movac}
\end{figure}

As presented in figure~\ref{fig:movac}, the band structure and  DOS  are presented for the  Mo-vacant system, the direct band gap  increases and  reveals new defect levels in the midgap at 0.35 eV, which is mainly composed of the hybridization from  Mo-4$d$ due to the antibonding states introduced by Mo-vacancy, and a shallow level at $-0.45$  from bonding due to the S-3$p$ orbitals around the Mo vacancy, then at 0 eV  (Fermi level) and highest peak value (19 states/eV) due to the contribution of S-3$p$ and Mo-4$d$ orbitals.  The downward shift of the Fermi level to the valence band  suggests that the Mo-vacant system is a $p$-type semiconductor. These results confirm previous studies~\cite{salehi2016atomic,santosh2014impact}.

\begin{figure}[!t]
	\centering
	\includegraphics[width=0.7\textwidth]{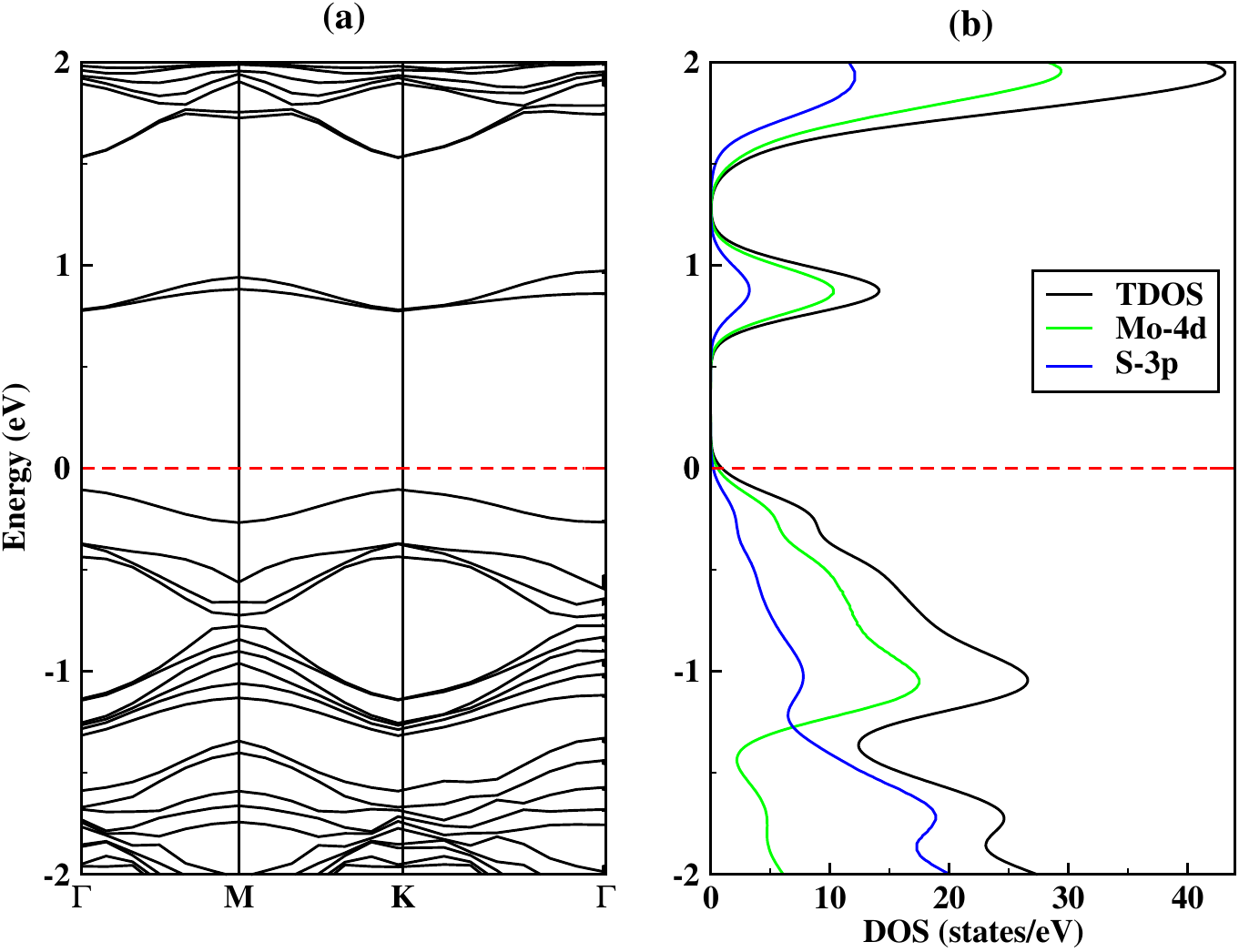}
	\caption{(Colour online) Band structure ($\mathbf{a}$), total and partial DOS of S-vacancy in monolayer  MoS$_2$  ($\mathbf{b}$). Fermi level is set at 0 eV.}
	\label{fig:svac}
\end{figure}

Figure~\ref{fig:svac} presents the band structure and  DOS of the S-vacant system. The presence of S vacancy increases the direct band gap and create localized defect levels, the first one located at 0.89 eV  arises from the contributions of Mo-4$d$ and S-3$p$ orbitals. Additionally, a shoulder near the valence band edge

\noindent located at $-0.24$ eV arises from the contributions of Mo-4$d$ and S-3$p$ orbitals. The Fermi level shifted downwards to the valence band suggests that the MoS$_2$ with  S vacant system is a $p$-type semiconductor~\cite{salehi2016atomic,zhou2013intrinsic,santosh2014impact}.
	
\begin{figure}[!t]
	\centering
	\includegraphics[width=0.7\textwidth]{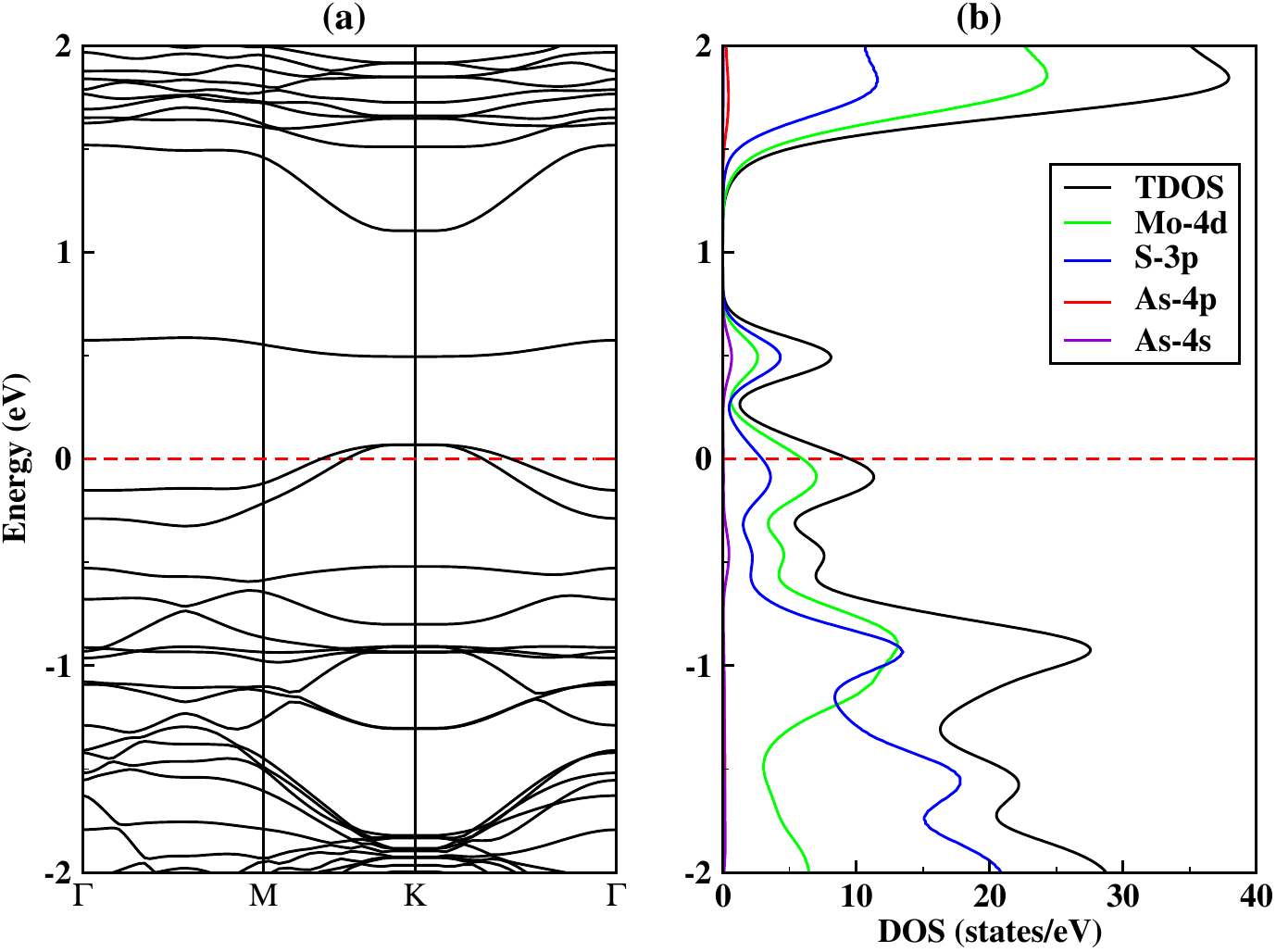}
	\caption{(Colour online) Band structure ($\mathbf{a}$), total and partial DOS of monolayer  MoS$_2$ with As doping at the Mo site  ($\mathbf{b}$). Fermi level is set at 0 eV.}
	\label{fig:dop}
\end{figure}

\begin{figure}[h]
	\centering
	\includegraphics[width=0.45\textwidth]{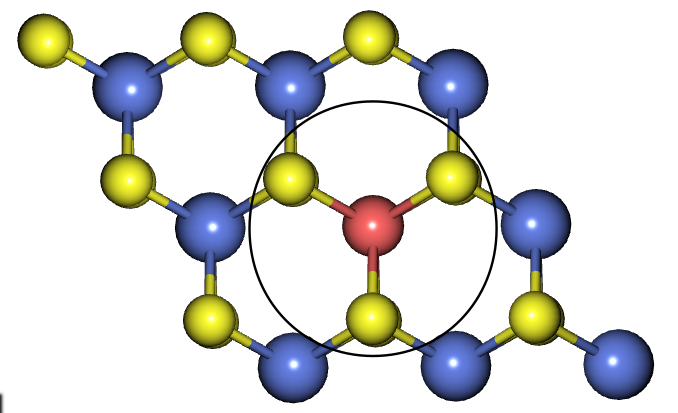}
	\includegraphics[width=0.45\textwidth]{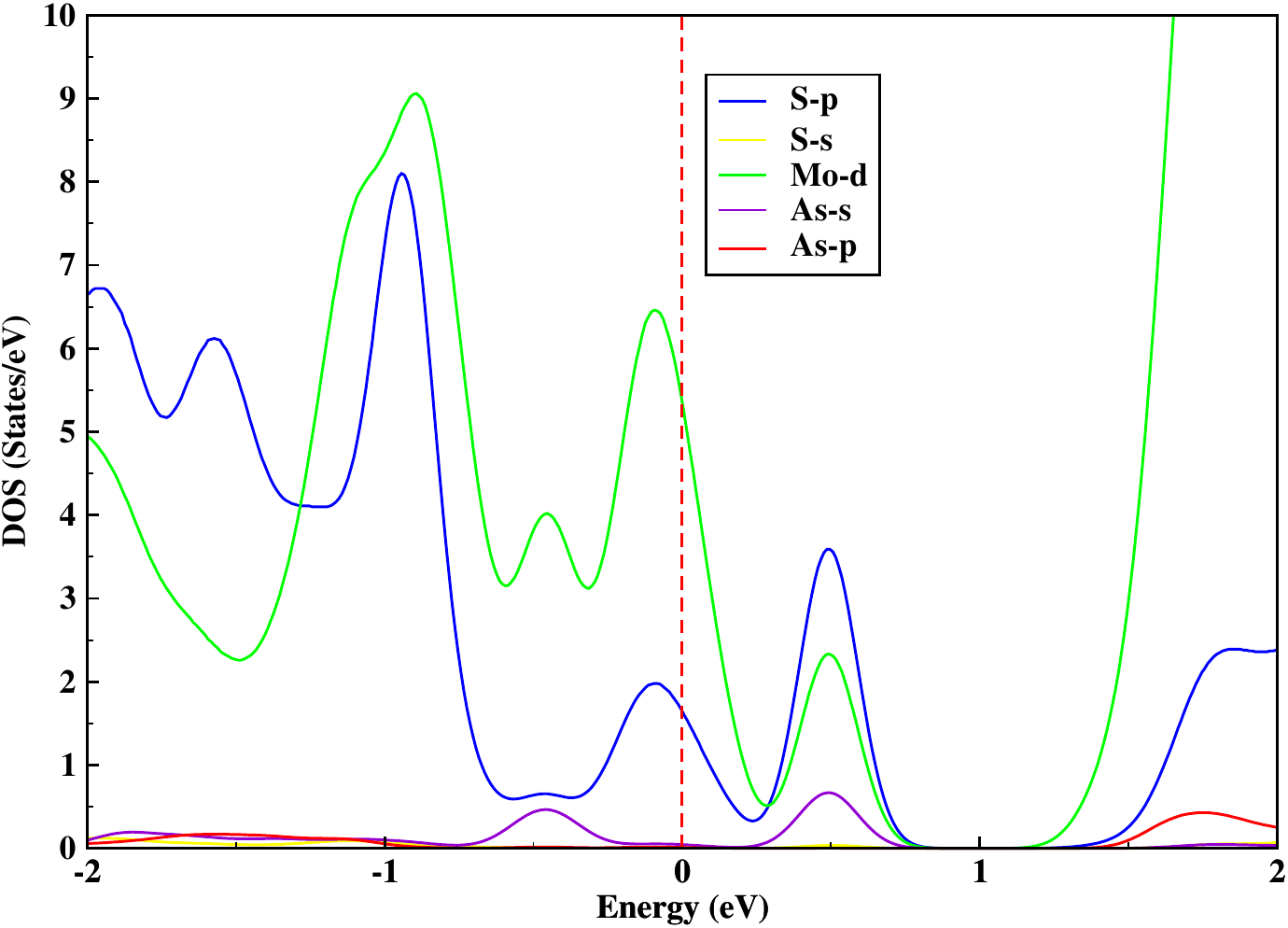}
	\caption{(Colour online) Geometric structures and PDOS of atoms S and Mo neighboring As:  As-Mo doped system. Fermi level is set at 0 eV.}
	\label{fig:geo and pdos1}
\end{figure}

	Figure~\ref{fig:dop} (a, b) illustrates the band structure and  DOS of MoS$_2$ monolayer where an As atom  substituted a  Mo atom in the vacancy position with concentration of 3.70\%.  In this case, the energy direct  band gap  has decreased  compared to the undoped MoS$_2$. From the figure~\ref{fig:dop} (b), the Mo-4$d$ and S-3$p$ orbitals predominately contribute to the valence band, with negligible contributions from the As-4$p$ and As-4$s$ orbitals. The Mo-4$d$ states predominantly contribute  to the conduction band and slight contributions from As orbitals. The PDOS for the As-Mo doped system reveals the emergence of two defect states located in the midgap. The highest level is at 0 eV (Fermi level), then the next appears about 0.5 eV  and a shallow level near to the valence band at $-0.45$ eV. The PDOS figure~\ref{fig:dop}(b) did not provide a clear  explanation of the origin of  defects, the contribution from the As atom was not observed due to the PDOS contained contributions from all the S and Mo atoms, owing to the presence of only one As atom. To gain a better understanding, we plotted PDOS for the S and Mo  atoms  neighboring the As atom, as presented in figure~\ref{fig:geo and pdos1}. We can see that the  defect level at 0 eV  (Fermi level) arises from the contributions of Mo-4$d$ and S-3$p$ orbitals  similar to the Mo-vacancy with a decrease in peak value (11.17 states/eV), although a contribution in the defect state at 0.5 eV (above Fermi level) is due to the S-3$p$, Mo-4$d$, and As-4$s$  orbitals, while the defect state located at $-0.45$ eV (below Fermi level)  arises from the contributions of Mo-4$d$ and S-3$p$  and As-4$s$ orbitals. The downward shift of the Fermi level to the valence band suggests that it is a $p$-type semiconductor. The density peak of As atom is smaller compared to Mo and S atoms. This indicates that the DOS for arsenic substituting Mo in monolayer MoS$_2$  is a combination of the effect of the individual contribution of the As atom and the Mo-vacancy. This insight is in agreement with both the exprimental and theoretical findings  for the Sb(Mo) and Nb(Mo) systems~\cite{zhong2019electronic,menezes2021unveiling,suh2014doping,dolui2013possible}. The presence of new defect levels and the decrease in the band gap plays an important role in photocatalysis~\cite{xie2020mo} and high-efficiency photovoltaics~\cite{beshir2022janus}.


\begin{figure}[!t]
	\centering
	\includegraphics[width=0.6\textwidth]{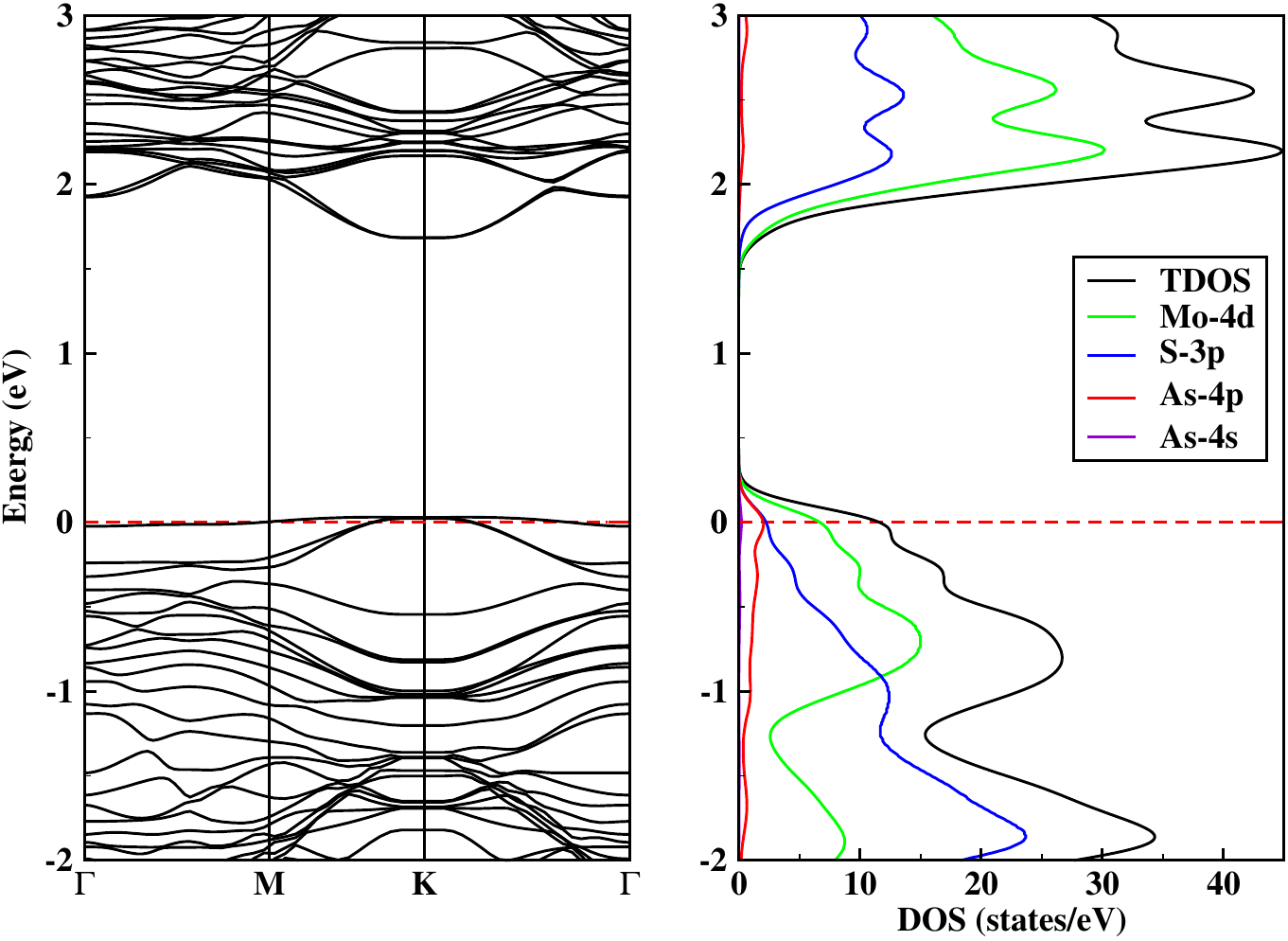}
	\caption{ Band structure ($\mathbf{a}$), total and partial DOS of monolayer  MoS$_2$ with As doping at the S site  ($\mathbf{b}$). Fermi level is set at 0 eV.}
	\label{fig:ass}
\end{figure}
	\begin{figure}[!t]
	\centering
	\includegraphics[width=0.45\textwidth]{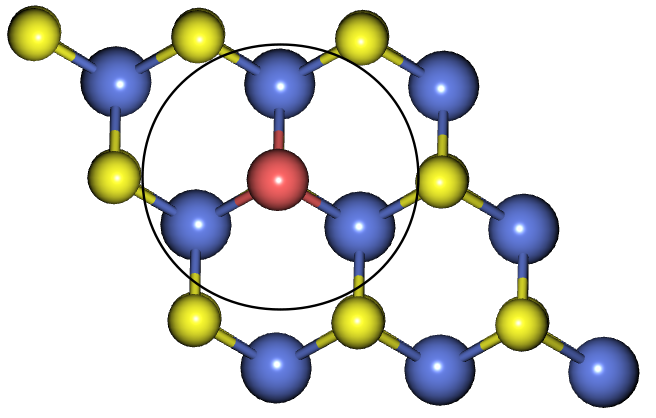}
	\includegraphics[width=0.45\textwidth]{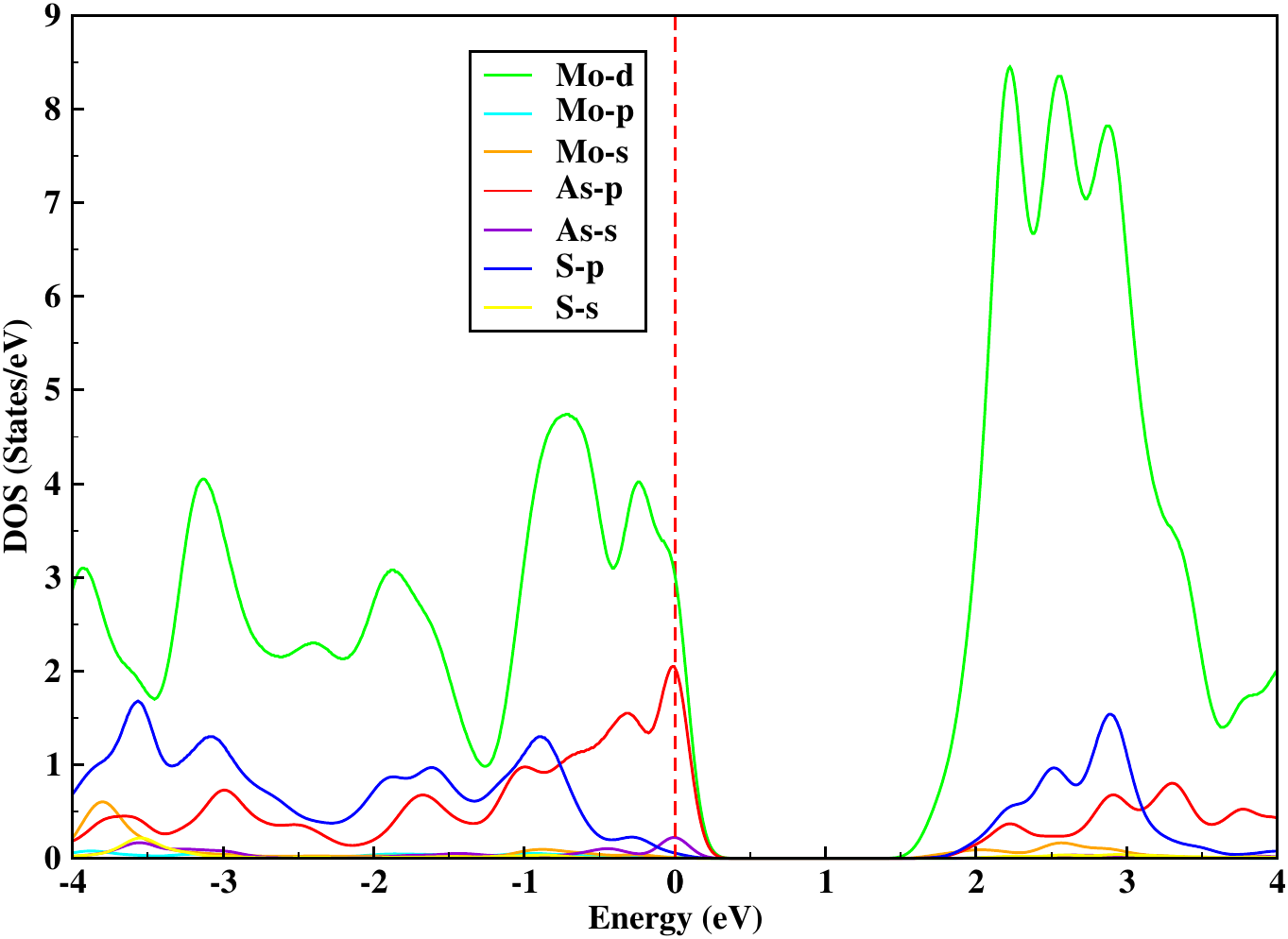}
		\caption{(Colour online) Geometric structures and PDOS of atoms S and Mo neighboring As:  As-S doped system. Fermi level is set at 0 eV.}
	\label{fig:geo and pdos2}
\end{figure}

	\begin{figure}[!t]
	\centering
	\includegraphics[width=0.7\textwidth]{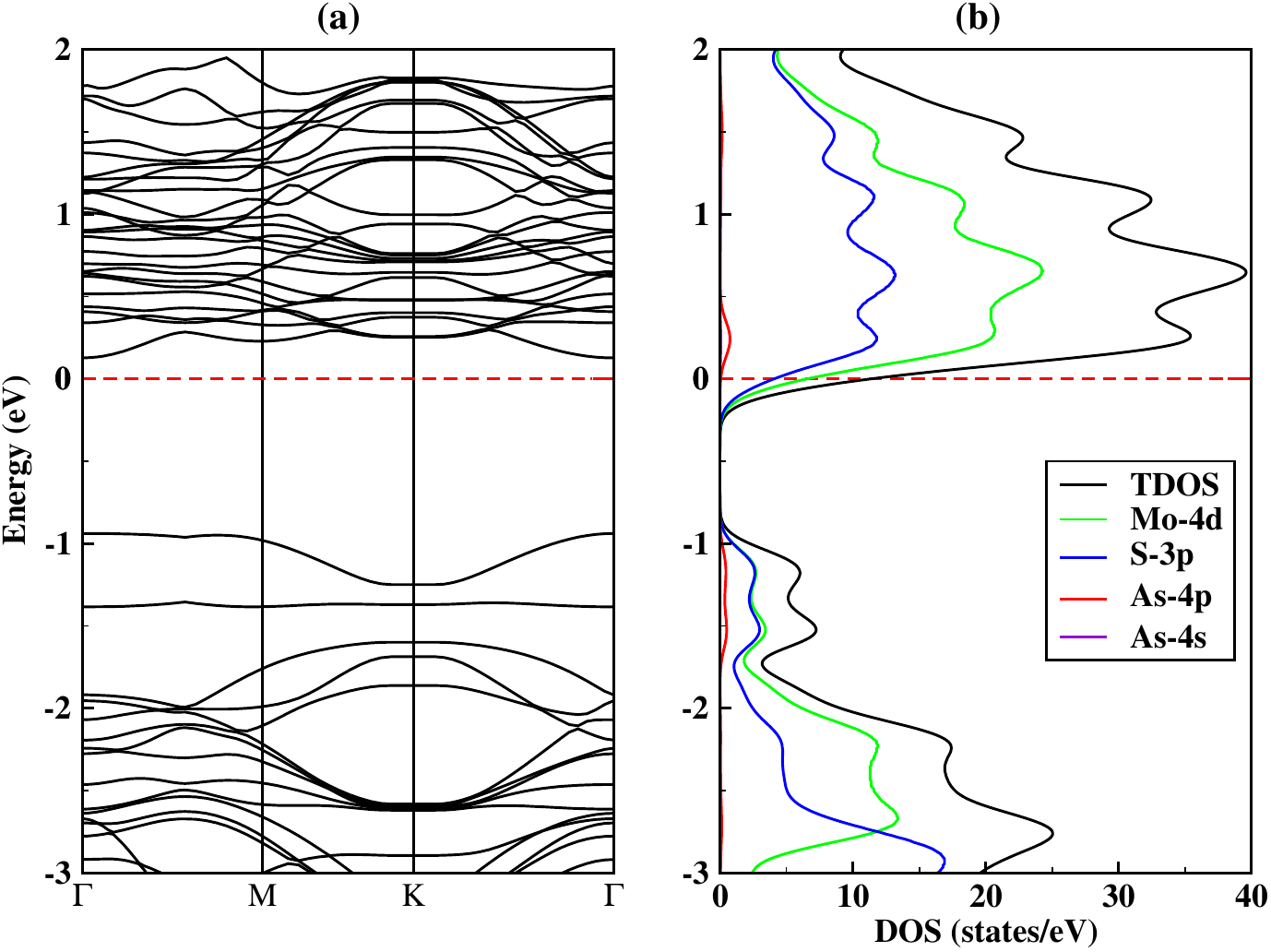}
	\caption{(Colour online) Band structure ($\mathbf{a}$), total and partial DOS of monolayer  MoS$_2$ with As interstitial doping  ($\mathbf{b}$). Fermi level is set at 0 eV.}
	\label{fig:inter}
\end{figure}
\begin{figure}[!t]
	\centering
	\includegraphics[width=0.45\textwidth]{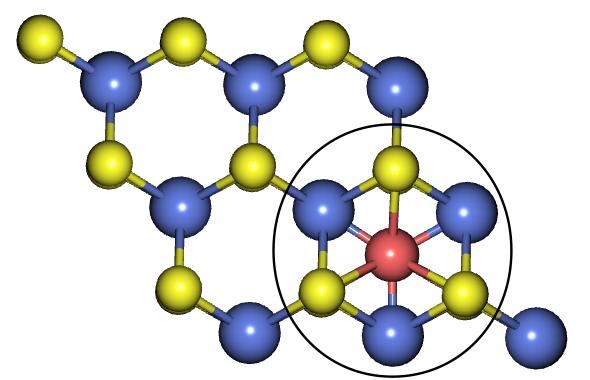}
	\includegraphics[width=0.45\textwidth]{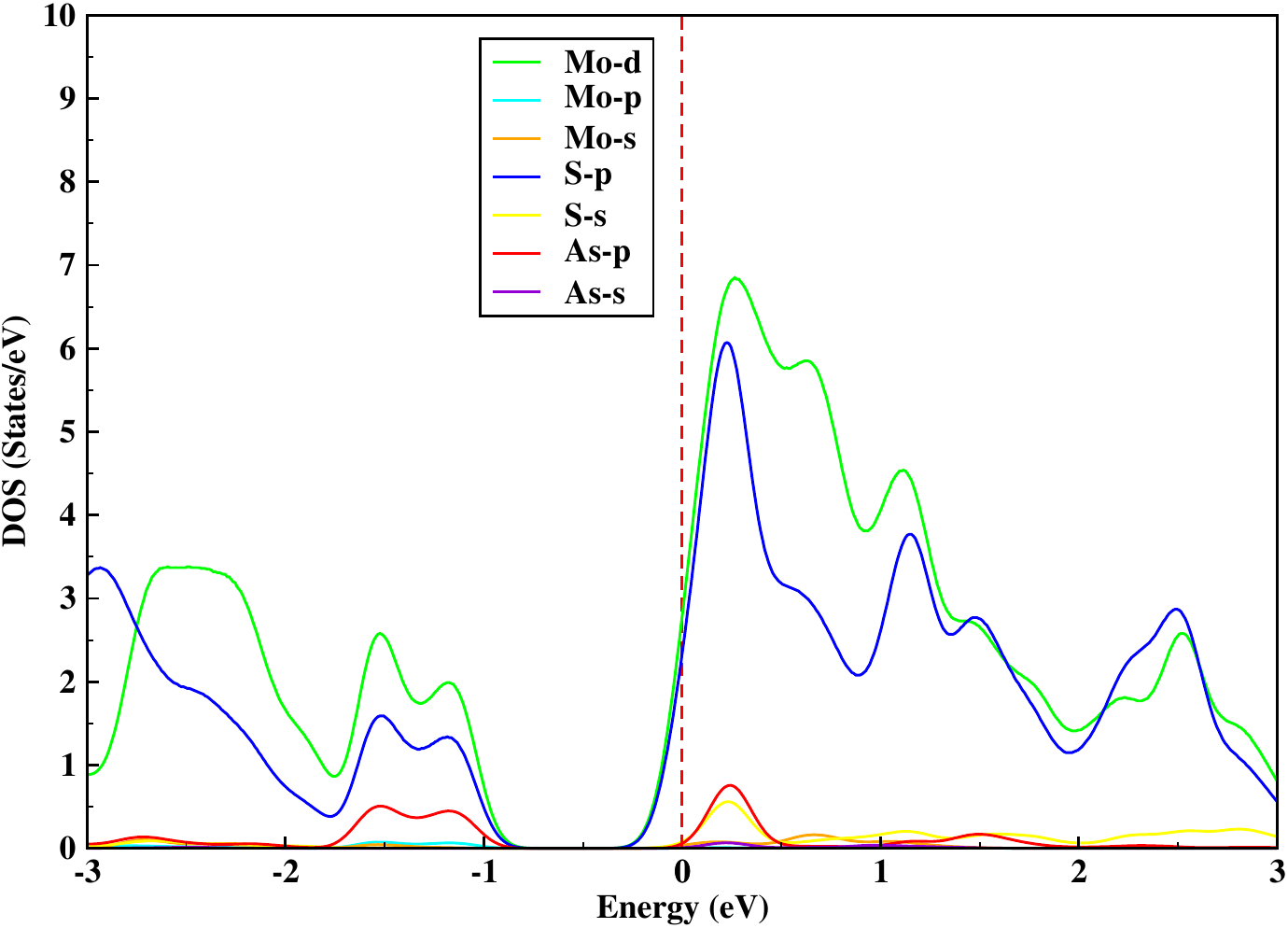}
		\caption{(Colour online) Geometric structures and PDOS of atoms S and Mo neighboring As: As interstitial doping system. Fermi level is set at 0 eV.}
	\label{fig:geo and pdos3}
\end{figure}
Figure~\ref{fig:ass} (a, b) presents the band structure and  DOS  of the MoS$_2$ monolayer, where an As atom  substitutes a  S atom at a vacancy site. Compared to the pristine MoS$_2$, the direct band gap decreased as detailed in table~\ref{t}. No clear and direct contribution from the As orbitals is observed in the PDOS. However, more significant changes in electronic density become more pronounced when analyzing the PDOS of the sulfur and molybdenum atoms neighboring the As atom (see figure~\ref{fig:geo and pdos2}), where a shoulder that appears attached to the valence band edge located at 0 eV, arises from the contribution of the Mo-4$d$ and As-4$p$ and As-4$s$ orbitals. Moreover, the shift of the Fermi level to the valence band suggests  that the doping introduces a $p$-type behavior. These results are consistent with the study by K. Dolui et al.~\cite{dolui2013possible}. The doping has been shown to improve  the photocatalytic efficiency of MoS$_2$ monolayer\cite{liu2024synergistic,mano2025tuning}.

In the system with an  interstitial  As atom positioned between Mo atomic plane and S atomic plane, the  figure~\ref{fig:inter} a and b display the band structure and the TDOS and PDOS.  It is observed that the energy  direct band gap  increased  compared to the undoped MoS$_2$ (see table~\ref{t}). The  TDOS and PDOS  analyses reveal two  defect levels in the midgap, at $-1.16$ eV and $-1.51$~eV. To determine the origin of these  defects, we also plotted the TDOS and PDOS for the interstitial As atom along with those of the neighboring Mo (2, 11, 14) and S (1, 3, 4, 6, 13, 15) atoms, as seen in figure~\ref{fig:geo and pdos3}. The PDOS indicate that these defect levels are due to the contributions of  Mo-4$d$, S-3$p$, and As-4$p$ orbitals. Additionally, the  shift of the Fermi level to the conduction band indicating arsenic atoms acts as a source of $n$-type doping, enhancing the performance in field effect transistors (FETs)~\cite{tong2015advances}.

\section{Conclusion}
	
In summary, we examined the crystal structures and electronic properties of undoped MoS$_2$ and its doped variants by  first-principles calculations, including Mo and S vacancies, arsenic substitution and interstitial system. The formation energy calculations showed that the As-Mo doped system has the minimum formation energy of 2.81 eV, indicating its higher stability, while the As interstial system has the maximum formation energy of 26.03 eV, making it the least stable and difficult to form.
Furthermore, we performed the analysis of electronic properties of the various doped systems comparing them with the undoped MoS$_2$. When Mo and S vacancies were introduced, the band gap increased and new defect levels appeard in the midgap. Specifically, for Mos$_2$ with Mo vacancy, there was exhibited a $p$-type behavior with defect levels  arising primarily from Mo-4$d$ and S-3$p$ hybridizations. The As-Mo doped system showed a further reduction in the band gap to 1.59 eV, suggesting its potential as a $p$-type semiconductor. The PDOS indicated that the defect levels located at 0 eV (Fermi level) because of the contributions of  S-3$p$ and Mo-4$d$  orbitals, with additional contributions from the As-4$s$ orbitals observed in the higher defect level at 0.5 eV. The results for the As-S doped system demonstrated a similar reduction in the band gap, but the contribution from As was less pronounced, indicating a less significant impact on the electronic structure compared to the As-Mo doped system. Overall, our results suggest that the doping of MoS$_2$ with As, especially at the Mo site, causes notable  modifications in the electronic properties. The alterations such as the presence of new defect levels and the decrease  in the band gap could enhance its performance in  photocatalytic and photovoltaic applications. The As interstitial system,  displayed $n$-type behavior with two defect level located in the midgap at $-1.16$ eV and $-1.51$ eV. These defects were attributed to contributions from Mo-4$d$, S-3$p$, and As-4$p$ orbitals. The upward shift of the Fermi level in the interstitial As system suggests its potential  in field-effect transistors (FETs).

\bibliographystyle{cmpj}
\bibliography{AssiaDAOUADI}
\newpage
\ukrainianpart
\title{Дослідження впливу легування арсеном на структурні та електронні властивості моношару MoS$_2$ з перших принципів}
\author{А. Дауді, М. Л. Бенхедір}
\address{Лабораторія теоретичної та прикладної фізики,  Ехаід Чейк Ларбі Університет Тебесси, 12000 Тебесса, Алжир
}
\makeukrtitle
\begin{abstract}
	Це дослідження спрямоване на вивчення структурних та електронних властивостей легованих моношарів MoS$_2$, включаючи вакансії Mo та S, а також леговані системи As з використанням розрахунків на основі методу функціоналу густини. Електронні властивості були проаналiзованi для з'ясування, як їх зміни впливають на поведінку матеріалу. Поява дефектів генерує нові дефектні стани в середині забороненої зони. У легованих системах з S-вакансіями (V$_\text{S}$), Mo-вакансіями (V$_{\text{Mo}}$), As-Mo (As, що заміщує Mo) та As-S (As, що заміщує S), зсув рівня Фермі вниз до валентної зони вказує на поведінку типу~$p$. У міжвузловій системі As рівень Фермі зміщується в зону провідності, що свідчить про напівпровідник $n$-типу. Результати показують, що легування MoS$_2$ As, зокрема на місці Mo, може бути використане у фотокаталізі та високоефективній фотоелектричній системі. Крім того, міжвузлова система As демонструє кращу продуктивність польових транзисторів.
	\keywords 2D матеріали, теорія функціоналу густини, квантовий пакет ESPRESSO, легування, електронна структура, дисульфід молібдену
\end{abstract}
\lastpage
\end{document}